\documentclass[preprint2]{aastex}

\def \eg           {{e.g.}}
\def \etal         {{et~al. }}

\def \ha           {\hbox{H$\alpha$}}
\def \kms          {\hbox{km$\,$s$^{-1}$}}

\def\approxlt{\lower.2em\hbox{$\buildrel < \over \sim$}}
\def\approxgt{\lower.2em\hbox{$\buildrel > \over \sim$}}

\def \ls           {\hbox{L$_{\odot}$}}
\def \ms           {\hbox{M$_{\odot}$}}           

\begin{document}

\title{STAR FORMATION ACROSS THE TAFFY BRIDGE: UGC~12914/15}

\author{Yu Gao\altaffilmark{1}\email{gao@astro.umass.edu}
Ming Zhu\altaffilmark{2, 3}\email{m.zhu@jach.hawaii.edu} 
E. R. Seaquist\altaffilmark{3}\email{seaquist@astro.utoronto.ca}}

\altaffiltext{1}{University of Massachusetts, Department of Astronomy,
LGRT-B 619E, 710 North Pleasant Street, Amherst, MA 01003-9305}
\altaffiltext{2}{Joint Astronomy Centre, National Research Council Canada,
660 N. A'ohoku place, University Park, Hilo, HI 96720}
\altaffiltext{3}{University of Toronto, Department of Astronomy and 
Astrophysics,  60 St. George Street, Toronto, ON, M5S 3H8,
Canada}

\begin{abstract}

We present BIMA two-field mosaic CO(1-0) images of the 
Taffy galaxies (UGC~12914/15), which show the distinct taffy-like 
radio continuum emission bridging the two spiral disks. 
Large amounts of molecular gas ($1.4\times 10^{10}\ms$, using the
standard Galactic CO-to-H$_2$ conversion applicable to Galactic 
disk giant molecular clouds [GMCs]) were clearly detected 
throughout the taffy bridge between the two galaxies, which, as 
in the more extreme case of HI, presumably results from a 
head-on collision between the two galaxies. The highest CO 
concentration between the two galaxies corresponds to the 
\ha~ source in the taffy bridge near the intruder galaxy
UGC~12915. This HII region is also associated with the strongest 
source of radio continuum in the bridge, and shows both 
morphological 
and kinematic connections to UGC~12915. The overall CO 
distribution of the entire system agrees well with that of the radio 
continuum emission, particularly in the taffy bridge. This
argues for the star formation 
origin of a significant portion of the radio continuum emission. 
Compared to the HI morphology and kinematics, which are strongly
distorted owing to the high-speed collision, CO better defines
the orbital geometry and impact parameter of the interaction, as
well as the disk properties (\eg, rotation, orientation) 
of the progenitor galaxies. Based on 
the 20cm-to-CO ratio maps, we conclude that the starburst sites
are primarily located in UGC~12915 and the \ha~ source in the bridge 
and show 
that the molecular gas in the taffy bridge is forming into stars 
with star formation efficiency comparable to that of the target 
galaxy UGC~12914 and similar to that in the Galactic disk. 

\end{abstract}


\section{INTRODUCTION}

Both observational evidence and theoretical perspectives show that
galaxy interactions and mergers are one of the most important
processes during the formation and evolution of galaxies. This is 
true both locally in the nearby universe and, in particular, 
cosmologically at high-redshifts. Interaction/merger-induced 
starbursts, nuclear
activity, feedback, and other associated phenomena, e.g., gas/dust
transformation, infalls and outflows, 
play a critical role in shaping the interstellar medium (ISM) and
intergalactic medium (IGM). Although almost all of the most luminous infrared
galaxies (LIGs) and ultraluminous infrared galaxies (ULIGs) 
are known to be involved in mergers of gas-rich systems 
(see Sanders \& Mirabel 1996 for a review), some
gas-rich interacting galaxies that are yet not very infrared-luminous or 
active in star formation still lack thorough investigation. Why the
interaction or merging in some gas-rich galaxy pairs does not
instantaneously lead to the luminous
starburst phase remains elusive.

The Taffy galaxies (UGC~12914/15, VV~254), comprise an extraordinarily
gas-rich pair with M(HI)$\sim 1.5\times10^{10}\ms$\footnote{In this
paper we use the following values $d_{\rm
L}=60$Mpc, $H_0=75$\kms~Mpc$^{-1}$} (Mirabel \&
Sanders 1988; Condon et al. 1993) and M(H$_2)\sim 4.5\times10^{10}\ms$
(this work, also Smith \& Struck 2001). The total molecular gas mass
in this system, derived using the standard CO-to-H$_2$ conversion
applicable to giant molecular clouds (GMCs) in the Galactic disk, is
about 3 times that of the archetypal galaxy mergers --- the Antennae
galaxies Arp~244 (Gao \etal 2001b; Zhu, Seaquist, \& Kuno 2003) 
and the nearest ULIG Arp~220.  But, strictly
speaking, the Taffy system is not a bona-fide LIG,
with only $L_{\rm IR}=8.1\times 10^{10}\ls$ (see Sanders \& Mirabel
1996 for the definition of $L_{\rm IR}$). 

Although Taffy is evidently a strongly interacting system, 
the global star formation efficiency, 
in terms of star formation rate ($L_{\rm IR}$) per unit of molecular 
gas mass, 
SFE$=L_{\rm IR}/M($H$_2) \sim 2\ls/\ms$ is rather low. This is nearly 
10 times lower than that of most LIGs (Gao \& Solomon 1999), which are 
mainly interacting gas-rich galaxy mergers. In this regard, Taffy 
is also similar to the Antennae
galaxies (Gao et al. 2001a) and a few known LIGs with lowest SFE, 
coldest far-IR colors, and abundant cold dust/gas, such as Arp~302 
(Gao 1996; Lo, Gao, \& Gruendl 1997), for example. How does the 
Taffy system retain 
globally most of the gas not actively forming into stars? Is the SFE 
really low or is the CO-to-H$_2$ conversion factor $X$ well off-scale, 
resulting in a serious overestimate of the molecular gas mass? Are there any 
starbursts going on at all? Where are the starburst sites? 

The Taffy pair is remarkable,
however, in that large amounts of radio continuum emission are found
between the galaxy disks, connecting the two galaxies in a taffy-like
structure (Condon \etal 1993). HI is also mostly distributed
between the two galaxies in the taffy bridge (Condon \etal 1993), 
rather than in the spiral disks or the tidal tails.
The Taffy not only shows an unusual HI morphology, but also the
HI kinematics indicate two fast counter-rotating disks (rotation 
speed near 300\kms, Condon et al. 1993) even after the collision. 
The highest HI column densities 
located in the taffy bridge are thus apparently 
a result of the stripped HI gas clouds owing to the direct cancellation 
of their motion after the counter-rotating, 
interpenetrating, and head-on disk collision.
It is conceivable that some GMCs were left behind 
in a similar way although the chance of direct 
collision between GMCs is rather small (e.g., Jog \& Solomon 1992).

While little emission is found in the bridge/gap region 
in the optical and near-infrared wavelengths, the H$\alpha$
image clearly reveals a distinctively separated H~II region 
near the disk of the northeastern galaxy UGC~12915
(Bushouse \& Werner 1990). 
This HII region is also the strongest radio continuum site
in the bridge. Jarrett et al. (1999) showed that there are several vigorous
star-forming regions in and near UGC~12915, based 
on the ISO mid-IR images, and one corresponds to this extra-disk
HII region. Condon et al. (1993)
also noted that synchrotron emission from relativistic
electrons trapped in the bridge nearly doubles the total radio
luminosity expected from spirals based on the tight far-IR/radio
correlation (e.g., Helou, Soifer, \& 
Rowan-Robinson 1985; Condon 1992). Then, the excess radio emission
in the bridge may have little to do with the enhanced star formation.
Is this indeed so? 

There is also a ring-like pattern in the southwestern (SW)
galaxy UGC~12914. The symmetric ring structure 
appears to be well explained by N-body simulations 
and is believed to be formed when
the ``intruder'' galaxy UGC~12915 goes through the rotating disk of the
``target'' galaxy UGC~12914 (Appleton \& Struck 1996; 
Gerber et al. 1990, 1996; Struck 1997). 
The well defined dynamical history of ring galaxies also makes the 
Taffy system an ideal candidate for studying the effect of galaxy 
interaction on the ISM and for understanding the interaction-induced 
star formation.
 
All these observations and numerical modeling 
are somewhat hampered by the
lack of knowledge of the detailed molecular gas distribution and kinematics,
which are essential to understanding star formation and constructing
the interaction history. Both the HI disk morphology and HI gas kinematics 
are too severely disturbed, and the spatial resolution is too low
to deduce the detailed galaxy disk kinematics of the pre-collision 
spiral pair. In comparison, molecular gas disks should suffer less damage in 
the early stage of galaxy interactions.
CO imaging can thus better probe the star-forming gas distributiion
and kinematics of the progenitor galaxies as well as current/future
sites of star formation. Our high-resolution CO observations 
and multiwavelength comparisons might provide a Rosetta stone to shed 
light on more detailed numerical modeling incorporating the multiphase 
ISM (e.g., Struck 1997; Barnes 2002).

\section{OBSERVATIONS}

The 10-element Berkeley-Illinois-Maryland Association (BIMA) 
interferometer (Welch et al. 1996) is the most
suitable millimeter instrument available at present to image the
extended CO distribution and kinematics of the Taffy
galaxies.  This is because the BIMA
has a large field-of-view of $\sim 2'$ at 3mm as well as short
baseline u-v coverage (the shortest baseline is $\sim 7$m in the
ultra-compact D array configuration). In addition, the
flexible correlator configuration covers a rather broad velocity
range. These characteristics are ideal for mapping extended CO emission 
of angular extent $\approxgt 40''$. 

Also because the entire Taffy system extends over 2$'$--3$'$, we adopted 
two-field mosaicing with each field-of-view's phase center 
pointing roughly near each galaxy's nucleus. This gives a sufficient 
overlap between the two BIMA pointings over the taffy bridge since the
nuclear separation is $67''$, much smaller than the BIMA's 
primary beam (FWHM$\sim 100''$) for CO(1-0). 

The observations were carried out during 1999 April---July. 
There were a total of 10 usable tracks accumulated for this project:
6 regular long (7--9 hrs) tracks at both the ultra-compact D and compact 
C array configurations, plus 4 short (4--6 hrs) tracks. 

The on-source integration was
equally split between the two fields after observing the nearby quasar 
0010+109 several minutes for phase calibration. Uranus was
also observed during each track in order to calibrate the antenna 
gains. Observations were carried out mostly under good/fair 
weather conditions. Details are summarized in Table~1.
Data reduction was performed mostly with Miriad. The maximum 
entropy method (MEM) was used only in deconvolution, although Sault et al. (1996) 
have implemented mosaicing algorithms which can use either
CLEAN or MEM for simultaneous deconvolution.


All calibrated u-v data for the 10 BIMA tracks have been combined and 
inverted with the mosaic mode, and two sets of channel maps were created with velocity resolutions of 20\kms~ 
and 40\kms. The cleaned beamsize is 
$9.''9\times 9.''7$, corresponding to natural weighting 
(robust=1.2) in order to maximize the signal-to-noise ratio.
We also produced a high resolution datacube of uniform-weighting
(robust=$-1.0$) with synthesized beam of $5.''0\times 4.''2$,
but throughout this paper we focus  on
the more sensitive low resolution datacube, because of their higher
sensitivity and dynamic range.
Further cleaning of the background noise was done in AIPS. 
Moment maps were then produced from
the cleaned datacube. For comparison with the VLA HI results 
(Condon et al. 1993), we smoothed our CO datacube 
to the same 18$''$ resolution by Gaussian convolution.

\section{RESULTS AND ANALYSIS}

\subsection{Molecular and Atomic Gas Distribution and Masses}

Figure~1 presents the velocity integrated CO intensity map (the 
moment zero map) in false color with the two 
overlapping BIMA fields. We now describe main features in 
the CO distribution 
associated with each galaxy -- namely the molecular gas disks/rings
and nuclear CO, the extra-disk CO concentration in the bridge
coincident with the HII region, and the very extended, 
rather diffuse molecular gas in the taffy bridge. Both galaxies show 
nuclear CO concentrations, but UGC~12915
(the intruder) exhibits the strongest nuclear CO
concentration. A significant quantity of CO in the bridge is 
associated with the HII region, which appears 
to be directly attached to the molecular disk of UGC~12915. 
This is much more clearly shown in the velocity channel maps
of 20\kms~ width presented in Fig.~2, which will be further 
detailed in the next section (\S3.2).
Moreover, the amount of molecular gas in the bridge between the galaxies 
appears to be comparable to that of UGC~12914 (target).



Fig.~3 presents the large-scale HI distribution of the Taffy
system (Condon et al. 1993), with respect to the optical structures. 
Obviously, the highest HI concentrations are in the bridge, 
and quite extended, rather than associated
with the disks.  Furthermore, nearly half of the total HI is 
located between the optical disks.  The white
contours in Fig.~3 show the location of the HI peak as well as the
half-maximum contour running through both optical disks. Unlike CO
which is mostly confined to the disks, HI is mostly located
in the taffy bridge and the tidal
tails, presumably the effects of ram pressure and tidal
stripping. In fact, the HI disk morphology of the individual galaxies has
mostly disappeared although kinematically recognizable, 
and the two HI disks seem to have already merged
into one system.


UGC~12914 has prominent stellar ring structures, which 
are also clearly visible in the DSS image (Figs.~2 \& 3). 
Unlike the luminous infrared ring galaxy NGC~1144,
where huge CO concentrations are distributed along the rings
(Gao et al. 1997), little molecular gas is detected 
in the ring of UGC~12914. However, the HI brightness might peak on 
part of the ring, and weaker HI
concentrations appear to follow the ring structures (Fig.~3),
although the limited spatial resolution doesn't allow
any detailed comparisons. Secondary CO concentrations are also
obvious in both galaxies, most likely indicating the locations
of the rotating CO rings/disks. 

We estimate the CO fluxes over different regions, as well as
the total CO flux of the entire system, using both
the velocity integrated CO intensity (moment zero) map (Fig.~1) 
and the velocity channel maps (Fig.~2). From optical images,
different regions such as disks, rings, and HII regions 
can be rather easily defined, although no distinct taffy feature 
is visible. The exact division of
different CO regions is not well defined, however. 
For instance, CO emission from the extra-disk
HII region is intimately connected to the CO disk of UGC~12915, 
which is mildly distorted. And it is 
difficult to assign the exact borderlines separating different
regions even in the velocity channel maps (Fig.~2). 
In particular, the separation of the HII region from
the diffuse taffy bridge CO is arbitary.
The adopted values based on our measurements over various regions are 
summarized in Table 2.


It is also difficult to distinguish the extended 
CO bridge from the CO possibly associated 
with part of the ring structures and the weak
CO near the strongest HI emission peak (Fig.~3), even using the CO
velocity channel maps. This is 
partly due to the limits imposed by the spatial resolution, particularly 
in the HI observations, the sensitivity
of our observations, and partly to the rather 
diffuse nature of the neutral gas (both molecular and atomic)
in the bridge. 
Also the CO fluxes can increase by up to $\sim 20\%$ if weaker extended 
regions of lower signal-to-noise ratio ($\approxlt 3\sigma$) 
are included.

Using the standard CO-to-H$_2$ conversion, 
$X \equiv N(\rm H_2)$/$I_{\rm CO}=3.0 \times 10^{20}$ 
$\rm cm^{-2}$ \rm (K km $\rm s^{-1})^{-1}$,
applicable to Milky Way disk
GMCs, we can estimate the molecular gas mass in different regions
across the system. We obtain the molecular gas mass directly from 
the estimated CO fluxes ($f_{\rm CO}=\int S_{\rm CO}dv$) in Table~2, 
using $M_{\rm H_2}(M_{\sun}) = 1.18 \times 10^4 D_L^2 f_{\rm CO}$,
where $D_L$ is luminosity distance in Mpc, which is equivalent 
to the standard CO-to-H$_2$ conversion.
Note the conversion from Jy/beam to Kelvin for the synthesized beam
of $9.''9 \times 9.''7$  is 0.99 K/(Jy/beam).
We thus obtain the total molecular gas masses of UGC~12914,
UGC~12915, the HII region, and the Taffy bridge to be 1.3, 1.5, 0.4,
and 1.4 $\times 10^{10} \ms$ respectively.

Several papers have reported single-dish CO observations of the
Taffy system. Smith \& Struck (2001) observed nine pointings using the
former NRAO 12m telescope and yielded five detections. 
They reported molecular 
gas masses of 1.9$\times 10^{10}\ms$ and 1.7$\times 10^{10}\ms$ for 
UGC~12915 and 12914 respectively. An upper limit of
$\approxlt 1.9\times 10^{10}\ms$ was also given for the taffy
bridge. Judging from the CO image, both of the 12m
beams pointing at UGC~12915 and at the taffy bridge completely cover
the dominant extra-disk CO feature. Thus the molecular gas mass
from the taffy bridge excluding the CO concentration at the HII region 
should be much less than this upper limit. Overall, a
total molecular gas mass of $\sim 4.5\times 10^{10}\ms$ for the whole
system seems to be in excellent agreement with our BIMA results.
 
Zhu et al. (1999) and Zhu (2001) also reported the total fluxes of the
Taffy system based on their observations with the NRAO 12m and IRAM
30m telescopes. Both measurements are in excellent agreement with our
BIMA results.  The IRAM observations (Braine et al. 2003) 
map the entire system at CO J=1--0 and 2--1 transitions with
11$''$ spacing. The comparison of the BIMA interferometer data and
the IRAM single dish data shows that the BIMA map has recovered
virtually all the CO flux within the calibration uncertainty of 20\%.  The
quantitative study of the excitation of the molecular gas in this
system combining the CO data at J=1--0, 2--1 and 3--2
transitions from different telecopes will be reported in a future 
paper (Zhu, Seaquist, \& Davoust 2003).

The Taffy system was also recently observed in the sub-mm 
continuum by Dunne et al. (2000). 
The 850$\mu$m fluxes of UGC~12915 and 12914 were listed as 160
and 131 mJy respectively. The total dust mass estimated is close to
10$^8 \ms$ assuming a single dust temperature, but including cold dust with
T$_{\rm cold}=20$K can increase the dust mass by a factor of 2
(L. Dunne 2003, private communication). Also, there may be significant
diffuse emission between the galaxies which is not yet detected.
Thus the actual dust mass can be 
much higher than the value reported above. 

The \ion{H}{1} gas mass (Table~2) can be estimated using
$M_{\rm H I} (M_{\sun}) = 2.36 \times 10^5 D_L^2 \int S_{\rm HI}dv $,
where $\int S_{\rm HI}dv$ is the \ion{H}{1} integrated flux in 
Jy km $\rm s^{-1}$. We use the estimated HI fluxes from 
the total HI line emission map shown in Fig.~3.

For a total gas mass (both HI and H$_2$) of 
6$\times 10^{10} \ms$,  we expect the total
dust mass to be on the order of 3$\times 10^{8} \ms$ assuming
a Galactic dust-to-gas mass ratio of 1/200. 

This value, 
based purely on the empirical gas-to-dust ratio, thus constrains
the CO-to-H$_2$ conversion factor $X$ (see \S4.3.1). Future more
sensitive sub-mm measurements including 450$\mu$m with SCUBA on the JCMT, 
as well as SIRTF 
far-IR observations, will firmly settle these issues.

\subsection{The Gas Kinematics}

In the velocity channel maps, rather detailed radial velocity 
information and the kinematic relationship of the various structures 
can be identified (Fig.~2).
The molecular gas in the SW galaxy (UGC~12914, the target)
first appears at $-360\kms$ (4040\kms)  in the inner
part of the hook-like structure near NW edge of the stellar disk.
With increasing velocity, CO emission shifts systematically towards 
SE  along the stellar disk, and eventually disappears 
around 280\kms (4680\kms). The much stronger 
CO in UGC 12915 (the intruder) first becomes prominent
at $-200\kms$ (4200\kms), but may  show a much 
fainter signature starting as low as $-360\kms$ (4040\kms). 
The rather strong
molecular gas concentrations in the intruder disk move with increasing velocity
towards the NW, following the expected rotation pattern. The molecular
gas in UGC~12915 shows a direct kinematic connection with the strongest 
CO concentration in the bridge, and disappears entirely
beyond 420\kms (4820\kms). The high velocity CO in the intruder
is associated with the dusty patches in the NW 
which are evident in the DSS image. This indicates the progenitor
disk of the intruder is slightly more tilted, relative to the 
target galaxy's disk, than that which appears in the DSS image.
Both the NW stellar extension and SE fainter 
edge are possibly related to the tidal features produced during 
collision. This is also evident by examining the 2MASS
images, which clearly show that the underlying stellar disk of the
intruder galaxy is confined to the dusty patches in the DSS image.
The sense of the rotation of the two molecular gas disks
is exactly opposite. Thus, both disks
have molecular gas in counter-rotation at speeds exceeding 
300\kms, even after the direct, interpenetrating, 
head-on collision.


The molecular gas in the taffy bridge has a much 
smaller velocity range than 
the velocity spread of the two counter-rotating molecular disks. 
Nevertheless, this material has
a velocity spread of over 200\kms. The emission emerges at $-60\kms$ (4340\kms),
becomes progressively more prominent around 80\kms
(4480\kms), and disappears at $\sim 200\kms$ (4600\kms). The velocity 
spread of the strongest CO feature in the bridge,  the HII region 
adajcent to the intruder galaxy, is identical to that of
the CO emission over the entire bridge region.
These features are perhaps better represented in the first moment 
(velocity field) and second moment (velocity linewidth) maps shown in Fig.~4.
These moment maps (Figs.~1 \& 4)
are derived from the detailed 
velocity channel maps (Fig.~2). 

In summary,  the CO kinematics 
and the overall CO morphology reveal two rapidly counterrotating CO disks
($\approxgt 300 \kms$) with a fairly large amount of 
extended CO emission in the bridge. The CO emission 
in the bridge, including the concentration at the
HII region, occupies a much smaller velocity spread than the
rest of the system (Fig.~4b), yet its mean velocity is about
the same as the systemic velocity of the whole.


Using the convolved CO datacube (to match the 
18$''$ resolution of the HI), and the VLA HI datacube, 
kindly provided by J. Condon, we constructed
the position-velocity (P-V) diagrams presented in Fig.~5a \& 5c.
For both galaxies,
the P-V maps show that the HI and CO are distributed roughly
in the same velocity range, but HI spans a much larger spatial 
extent. In particular, for both 
CO and HI along the major axes, the P-V 
plots clearly show the rotation pattern of 
the gas disks. 

To reveal more clearly the difference between the HI and CO gas properties 
in the bridge, we also show an additional P-V diagram 
across the bridge (Fig.~5b). The P-axis joins the nuclei 
of the two galaxies, and passes through the CO concentration in 
the HII region. 
Clearly, there are two velocity components in HI
at $\sim$4500 and $\sim 4250 \kms$, whereas almost the entire 
CO emission in the bridge has only the $\sim$4500 \kms component.
In addition, a clear kinematic connection between the intruder
UGC~12915 and the CO in the bridge is evident from this P-V
plot as the cut is roughly along the minor axis, goes through
HII region and the bridge. 

Evidently, CO in both galaxy disks retains the signature
of the solid-body rotation pattern, whereas the HI disks 
are certainly distorted and/or lopsided 
(Fig.~5a \& c). In both HI gas disks, the distortion/lopsidedness 
is associated with the higher velocity parts. 
In fact, the extended HI
emission appears to be reminiscent of the long HI tidal tails observed
in most galaxy mergers. In extreme cases like the Antennae galaxies, 
more than half of HI appears to end up in the HI tidal tails 
(Hibbard et al. 2001a). 

The foregoing comparisons indicate that both CO disks have mostly survived the head-on
collision, and retained $\approxgt 300 \kms$ rotation velocity
(Fig.~5a \& c), whereas the HI disks have been disrupted, exhibiting
smaller rotation velocity than that of CO, and possessing a very asymmetric
distribution, both spatially and kinematically, owing to the strong
disk collision.

In Fig.~6, the grid spectra of both CO and 
HI in the bridge gap also show a distinct difference: one peak for 
CO but a double-peak for HI. Obviously, CO is associated with 
the higher velocity HI component in the bridge. 

From the HI channel map shown in 
Fig.~5 of Condon et al. (1993) and the overall HI morphology 
(Fig. 3), it is also evident that both of the HI disk extensions 
and the HI tidal tails (towards SE of the target and NW of the 
intruder) are in the higher velocity ranges. But, the severe 
damages in the HI disks are more associated with 
the lower velocity HI, which is now mostly in the taffy bridge.
This is also clearly shown in the 
P-V plots (Fig.~5), particularly in the target galaxy UGC~12914,
where most of lower velocity HI has been 
stripped away and left behind in the bridge.


From the spatial distribution of both CO and HI in the bridge, 
it may be seen that there is almost no CO detected coincident with the HI 
peak position (Figs.~1--3).
This is  clearly revealed from the CO (Fig.~2) and HI (Fig. 5 in 
Condon et al. 1993) channel maps over roughly same velocity 
range in the bridge. 
Most of CO in the bridge is in the HII region, and to the south of 
the HI peak, which might be related to the ring structures
of the target galaxy. This is partly reflected in the P-V
plot cutting through the bridge (Fig.~5b), as well as the spectra
of HI and CO near HI peak (Fig.~6). Except for the
HI extensions/tails associated with the tidal features in the
galaxy disks, most of 
HI is rather extensively distributed across the bridge
in two distinct velocity ranges, and possibly originates
from both galaxies.

\subsection{Star Formation Efficiency Map}

The ratio of $L_{\rm IR}/M({\rm H}_2) (\sim 2~\ls/\ms$ for the 
Taffy system) is often
referred as the global star formation efficiency (SFE) since the 
total FIR emission, the tracer of the global star formation rate,
has been normalized by the total molecular gas mass available 
to make stars. Similarly, we can quantify the local star 
formation properties by measuring the local SFE. We have 
localized measurements of the molecular gas mass from CO maps. 
We here characterize the star formation rate simply 
by using the radio continuum emission as a surrogate for the FIR 
emission (see \S~4.2; also Murgia et al. 2002). We thus 
obtain the radio-to-CO ratio map to reveal the variation in local SFE. 
A successful application of this technique to the Antennae galaxies
can be found in Gao et al. (2001a). 
The VLA 20cm continuum (Condon et al. 1993) and our CO maps 
have roughly comparable spatial resolution. 
Thus a direct division can be obtained once the 
20cm continuum map has been regridded to match the pixel
scale of the CO map.


Remarkably, the local 20cm/CO ratio (Fig.~7) changes in most cases only
by factors of a few and stays roughly constant over the bulk of the
molecular gas distribution. The lowest contour plotted 
is the geometric mean of the minimum and maximum of the 20cm/CO 
ratio.  
Successive
contours increase by a factor of $\sqrt {2}$.  The peak
contour is only a factor of 5 larger than the average.  The
highest SFE sites, with SFE$\approxgt 11~\ls/\ms$, are in the 
nuclear region and in the inner edges of the
molecular gas ring/disk of UGC~12915. The adjacent extra-disk 
starburst site (the HII region) in the bridge has 
SFE$\approxgt 7~\ls/\ms$, nearly twice that of the Galactic 
GMCs. The inferred SFE across most
of the taffy bridge is comparable to the highest found
in the nuclear region of the target galaxy UGC~12914, 
which is about the same as
that of the Galactic GMCs. The region around the highest HI
peak, where little CO emission is found, has slightly higher SFE
than even the nuclear region of UGC~12914.

Therefore, the sites of highest
SFE are exclusively associated with the possible starburst sites 
in the intruder. Other plausible active star-forming sites of 
modest SFE are perhaps in the taffy bridge, one at the HII 
region, and the other near the highest HI column density. 
The other regions in the taffy bridge as well as the target galaxy 
have normal SFE comparable to that of Galactic GMCs, and
are thus just normal star-forming sites.

\subsection{The Multiwavelength Comparison}

Fig.~8 presents near-IR H-band, ISO mid-IR 15$\mu$m 
(Jarrett et al. 1999), VLA 20cm continuum, and HI line emission
(Condon et al. 1993), overlaid with CO contours. Although it
is quite faint, the HII region also shows up 
in the near-IR image (Fig.~8a). Even 2MASS images appear to indicate
some faint emission at the HII region. This HII region is also
evidently present in the  15$\mu$m image (Fig.~8b), although it is 
not certain whether other mid-IR features in the taffy bridge are 
real or not. 

The best match in Fig.~8 is between the radio continuum and the CO 
emission, including the taffy bridge. Most of the main features,
either in the disks or in the taffy bridge, 
appear in both 20cm and CO images 
(Fig.~8c). However, it is noticeable that 
the northern portion of the taffy bridge, evident 
in radio continuum, is rather deficient in CO emission. 
The deficiency in CO in some location is likely real, in particular,
some small gaps appearing in the SFE contour map (Fig.~7) due to 
the lack of CO emission there.
Comparison between CO and HI (Fig.~8d) indicates that 
most of the extended HI concentrations are 
systematically shifted towards the NW direction relative to 
most CO emission in the bridge.


With all observations combined, in particular, the CO and HI 
kinematics, we may better constrain the various parameters
of the two galaxies. Given the strong distortion of the
HI morphology/kinematics, parameters like the maximum corrected 
rotation velocity $V_{\rm max}$
at radius $R_{\rm max}$ are better deduced from CO data. 
This is because the molecular gas disks have suffered less 
disruption than the HI disks. 
Furthermore, the inclination of UGC~12915, 
determined from the near-IR morphology, is slightly less 
than that apparent in the optical image (Fig.~6a). The tidal 
feature elongated towards the NW leads to slightly higher
estimate of the inclination. Compared with published values (Giovanelli \etal 1986) we use an inclination of 
$\sim 70^\circ$ rather than 73$^\circ$ for UGC~12915, and 61$^\circ$ 
for UGC~12914, identical to the previously published value. 

The observed CO velocity linewidths are $W=600\kms$ 
and 580\kms, for UGC~12915 and UGC~12914 respectively, and thus 
$V_{\rm max}=315\kms$ and 327\kms~ after the inclination and
redshift broadening corrections. In comparison, the HI linewidths 
of 510\kms~ and 550\kms~, $V_{\rm max}=260\kms$ and 310\kms~ 
were given for UGC~12915 and UGC~12914 
respectively by Condon et al. (1993). Note that the value of 
$R_{\rm max}$ estimated from HI data might be too large 
owing to the contamination by the HI tidal features. 
Therefore, we use CO data which give $R_{\rm max}=0.41'$ 
and $0.25'$ to estimate the dynamical 
masses of UGC~12914 and UGC~12915, which are 
4.5 and 2.6$\times 10^{11}\ms$ respectively. 

\section{DISCUSSION}

\subsection{The Gas Pile-up and Dynamics in the Taffy Bridge}

When two counter-rotating gas disks collide at 
high speed, the diffuse gas clouds 
(mainly HI) collide at highly supersonic speeds. 
Thus, we can expect an extensive ``gas splash'', and possibly 
other observable effects of shock heating and radiative cooling 
as probed by numerical simulations (Struck 1997; Barnes 2002). 
The Taffy system is particularly unique, with a nearly 
face-on collision, and the 
relative collision speed of HI clouds is nearly 900\kms~, 
since the estimated transverse collision speed of the two galaxies 
is about 600\kms (Condon et al. 1993). The velocities of the HI clouds
could be essentially canceled after a 
direct, counter-rotating, and dissipative head-on collision.
Apparently, the molecular gas disks have mostly survived 
the collision. The HII region, however, may have been recently formed 
out of the extra-disk molecular gas concentration that has 
been pulled out of the molecular gas disks and condensed 
after the collision. 

A simple estimate can be made for the amount of HI left in the 
taffy bridge. For a perfect 
face-on disk collision with same counter-rotating
speed, all HI clouds in two galaxies that have collided 
inelastically  will be left behind after collision. 
Essentially almost all diffuse gas could be pulled out of the 
disks when the collision velocity is much smaller than the 
rotation speed. When the collision speed is much 
larger than the rotation speed, however, the amount of gas that 
collides will be less, depending upon the exact 
area of the disk-to-disk overlap and the amount of  area of the
disks that has been swept owing to the counter-rotation 
during their close impact. The Taffy system 
falls between the two
extreme cases as the collision speed is comparable to the 
counter-rotating speed.

Judging from the multi-wavelength images, particularly 
the radio continuum and HI maps compared to CO (Fig.~8), 
the intruder may have collided
with the inner nuclear region and the entire NW portion of the 
target disk with an angle of nearly 30$^{\circ}$ between the major 
axes. According to Toomre \& Toomre (1972), the impact parameter 
should be smaller than the disk radii in order to produce 
the ring structures, rather than the prominent 
tidal tails. However, there are weak tidal features 
present. Thus the impact parameter 
can't be much smaller than the disk radii either, since
numerical simulations show that encounters with impact parameter
larger than the galactic radii can be most effective in 
producing long tails and stronger gas inflow toward the nuclei.
Given these considerations, most of the SE portion 
of the target disk might be much less damaged. A small 
portion of the intruder's HI disk in the NW may have 
suffered less impact as well, owing to this particular 
impact configuration. Judging from the HI distribution in
Fig.~3,  we can assume that
more than 1/3 of HI clouds in the target galaxy reside in the
SE disk and about 1/3 of HI gas in the intruder galaxy is 
located in its NW disk. Then most of HI in the intruder
and the HI in the NW portion of the target galaxy, 
totalling about half of the total HI gas of the system, 
could have experienced a collision,
owing to the sweeping of the counter-rotating disks, 
with the remnants mostly left behind 
in the taffy bridge. 
A direct comparison of the HI morphology with the CO image
and the stellar disks (Fig.~8) appears to indicate that indeed 
nearly $\sim 50\%$ of HI is located in the taffy bridge 
between the two disk galaxies.

Our BIMA CO maps
obviously reveal abundant H$_2$ gas in the bridge.
In particular, there is a high concentration of molecular gas
near the intruder, coincident with the \ha~ source, indicating
clearly an active extra-disk star-forming region.
The spatial distribution and kinematics of CO along the bridge 
suggest that most CO is, in fact, the gas splash following 
the interpenetration of two inner molecular disks. 
Interestingly, except for the ring structures and the
molecular gas concentration in the HII region, little on-going 
star formation is detected in the bridge. The dust extinction,
resolution, and sensitivity could all be the contributing factors
for detecting few stars. Although
the peak HI column density at the taffy bridge is only 
3$\times 10^{21}$cm$^{-2}$, as measured by VLA 18$''$ beam, 
the CO emission observed by the BIMA $10''$ beam 
in the bridge actually
dominates the column densities (using the standard CO-to-H$_2$
conversion factor). The lowest column densities 
in the bridge detected in CO are in fact
several times larger than the peak HI column density. 
The estimated extinction
near the HI peak region, where the CO is weakest,
could still be quite high, with $A_V \approxgt 10$ magnitudes. 

Besides the obvious starburst site of the HII region in the bridge, 
the huge amount of HI piled up between the two galaxies could
be the potential sites for the development of the overlapping starbursts 
which are quite common in interacting galaxies (Xu et al. 2000).
Such large HI gas concentrations are also found in galaxies
of different interaction geometry, e.g., NGC~6670 which involves
an edge-on collision (Wang et al. 2001). It is conceivable
that once the HI concentrations further condense, and perhaps form
into self-gravitating systems, they could capture some GMCs passing
by, and could also possibly convert some HI into molecular form. 
An extreme example with prominent overlapping starbursts is 
the Antennae galaxies, where more than 50\% of the total CO emission is
found in the overlap region (Gao et al. 2001a). In the Taffy system, 
the huge HI concentration 
between galaxies might then provide a mechanism for an eventual 
molecular concentration in the bridge region.

High speed (900\kms) gas cloud collisions should produce
shockwaves heating the gas to more than 10$^7$K. It would be
interesting to map the hot ISM of the taffy bridge and further
compare the gaseous structures of different phases of ISM.
The Taffy system appears to be an intermediate example, 
bridging the slow-speed ($\sim 300\kms$ or less) gas-rich merging 
galaxies and the very high-speed ($\sim 1000\kms$) encounters 
in compact groups of
galaxies. In most merging systems, HI appears to roughly 
follow the distribution of the stars during the encounter,
whereas huge HI reservoirs, located far away and
completely outside of the galaxy disks, can often be found in
compact groups of galaxies (e.g., HCG~92, Williams, Yun \& 
Verdes-Montenegro 2002).
Also in HCG 92, CO is found in the surrounding IGM HI concentrations
(Gao \& Xu 2000; Lisenfeld \etal 2002). Prominent
large-scale \ha~ and soft X-ray emission, presumably related to 
the HI splashes and shockwave heating, are also 
evident in HCG 92 (Trinchieri \etal 2003).

\subsection{Star Formation across the
Taffy Bridge between the Galaxies}

It is tempting to interpret the 20cm-to-CO ratio as an SFE map (Fig.~7). 
This is because the radio continuum 
emission can be used as a tracer of the recent star formation in 
galaxies (\eg, Condon \etal 1996;
Condon 1992). In particular, there is an excellent correlation between 
the FIR and radio continuum emission (Helou et al. 1985; 
Condon 1992; Xu \etal 1994; Yun, Reddy, \& Condon 2001), and the tight 
FIR/radio correspondences appear also to be valid at least 
on kpc scales in galaxies (\eg, Marsh \& Helou 1995; Lu \etal 1996).
Murgia et al. (2002) 
analyzed the 20cm/CO ratio in a large sample of star-forming
galaxies and showed that it remains constant within a factor 
of 3 (with a resolution of about 50$''$), both inside the 
galaxies and from galaxies to galaxies.
Nevertheless, it is still debatable whether the radio continuum emission
in the taffy bridge is a good indicator of star formation.
Condon et al. (1993) noted that the synchrotron 
emission from relativistic
electrons trapped in the bridge nearly doubles the total radio
luminosity expected from normal spirals. Thus it is possible 
that the star formation rate estimated
from the radio continuum could be overestimated at most by a factor of 
two since synchrotron emission from relativistic electrons accelerated in
gas collision shocks may contribute
a significant fraction of the radio continuum. The close
correlation between CO and radio continuum, particularly in the extended
taffy bridge (Fig.~8c), however, may imply that most of the radio 
continuum emission could still be 
related to the very recent on-going star formation (e.g., 
Murgia et al. 2002). The radio continuum 
might be directly associated with the various by-products of the 
star formation process, such as supernovae remnants, and is therefore 
still a useful indicator of the (far-IR) star formation. 

The gas (HI + H$_2$) surface density in the bridge is significantly 
higher than the threshold for star formation as suggested by 
the spatially resolved study of normal spirals (Martin \& Kennicutt 2001). 
The HI gas alone in the bridge, as mapped by the $18''$ VLA 
beam, is about 20~\ms~pc$^{-2}$ (the second white contour 
outlining the HI gas between the two
galaxies in Fig.~3). The H$_2$ gas surface density is much 
higher, and the minimum detected by the $9.''9 \times 9.''7$ beam, 
using a standard conversion factor $X$, is as large as 
150~\ms pc$^{-2}$. Therefore, essentially all regions with CO
detection have molecular gas surface density about one order of magnitude
above the mean star formation threshold in normal spiral galaxies.
It should be noted that the HI peak region could be the
overlap of two HI arms (one arm from UGC~12915 and the other from 
UGC~12914), given the two distinctively different velocities 
in the bridge (Fig.~6). The actual HI column density at the 
HI peak regions could thus be a factor of two lower. However, the combination of 
the HI and H$_2$ (even with a conversion factor $X$ ten times 
smaller) still yields a gas surface density higher than the threshold
for star formation in normal galactic disks.

\subsection {Is the Taffy System in a Starburst Phase?}

Star formation occurs within 
GMCs, especially the dense cores where stars are forming with SFE  
orders of magnitude higher than the average. 
Interestingly, local starburst galaxies have SFE higher 
(by factors of 2--5) than normal spiral galaxies, whereas 
almost all LIGs/ULIGs have SFE 
$\approxgt ~20~\ls/\ms$, or even orders of magnitude higher
than normal spirals (Sanders \& Mirabel 1996). However,
there are some early stage pre-merging LIGs with SFE $\sim 10 ~\ls/\ms$
and lower, such as Arp~302 (Gao 1996; Lo et al. 1997).
Across most of the active star-forming regions in the Taffy system, however, 
starbursts are not particularly strong. Apart from 
the nuclear region of the intruder and its edge-brightened inner
CO disk/ring, the HII region in the bridge
barely qualifies as a true starburst site simply based on
its SFE. Since the average SFE of
the Galactic disk GMCs is a factor of 2 higher than the 
global SFE of the Taffy system, the entire 
system seems not to be a starburst at all.
Unless we have seriously overestimated the molecular gas mass 
by using the standard CO-to-H$_2$ conversion, almost all molecular gas
in taffy bridge and in the target galaxy has low SFE, and is
not in the starburst phase. 

The same arguement may apply to both the intruder and the
target galaxies as well since much smaller molecular gas mass
will result in extremely high SFE, which appears 
also to be inappropriate. In particular, because the Taffy system 
is itself not a genuine LIG system yet. 
The tight correlation between radio continuum 
and CO may actually indicate that the bulk of radio continuum
emission in the taffy bridge is probably related to the globally
normal star formation. And it is the large scale star formation 
in addition to some localized modest starbursts that
power most of the energy output in the Taffy system.

\subsubsection{Effect of the CO-to-H$_2$ Conversion Factor}

Any reduction of the conversion 
factor $X$ in the Taffy system will result in an enhanced SFE by 
the same factor. If we sufficiently overestimate the molecular gas mass in
the intruder, then the intruder could
clearly be a strong starburst. Similarly, if a smaller
conversion factor $X$ is used in the bridge, the molecular gas
there could then be forming into stars with
elevated SFE as well, even after lowering the star formation 
rate by a factor of 2 to bring it in line with the
FIR/radio correlation. It is important to note in this context that 
the conversion factor $X$ in ULIGs is significantly lower than 
in Galactic GMC's. In a ULIG, however, extremely
high nuclear gas concentrations around the merged double-nucleus
have resulted in such extraordinarily high molecular gas density
that even the intercloud medium is molecular. Thus,
most of CO detected in ULIGs is not from self-gravitating GMCs. 
Nevertheless, even under such extreme conditions, the
reduction in the conversion factor $X$ is only a factor of
$\sim 5$ (Downes \& Solomon 1998). Judging from the comparatively 
lower CO concentrations in the target galaxy UGC~12914, it is unlikely 
that such an overestimation of molecular gas occurs in this galaxy. 
However, this could be the case in the intruder galaxy where CO is
rather highly concentrated. 

However, given the mostly low concentrations of CO and lack of 
independent evidence for high 
star formation activity, it is unlikely 
that we have overestimated the molecular gas mass by
large factor, and we conclude that star formation is probably 
rather inactive over
bulk of the molecular gas in the Taffy system.

\subsection{The Taffy Galaxies as a Future Starburst System}

Like the Taffy system, many interacting systems show little or no 
enhanced new star formation, while statistically as a whole, 
they only show a factor of 2 enhancement
compared to normal spirals (Xu \& Sulentic 1991). Half of an
objectively selected sample of 8 interacting systems imaged 
by ISOCAM show no particularly obvious starbursts
(Xu et al. 2000). It is not surprising that high-speed 
splashes and shockwave heating may inhibit star formation 
since they disrupt disk clouds, and lead to an overall rarefaction 
of disk gas, and thus only some regions that have built up
enough gas concentrations can actually be actively forming into
stars, and are thus in a starburst phase. 

Yet it seems clear that interaction as well 
as the associated gas-infall and high-pressure 
compression of GMCs (\eg, Jog \& Solomon 1992) are important 
for triggering enhanced star formation, though it may only be 
preferred in mergers with a relatively slow-speed encounter. In any case, it
appears that such high-pressure compression might only be the initiator 
of a multi-step process. The Taffy system could have just experienced 
a strong starburst phase a few tens million years ago 
($S_{\rm nuclear~ separation}/V_{\rm transverse} \sim 3\times 10^7$yrs)
at the closest splash of the two disks. We could perhaps also expect 
a future starburst once the two galaxy disks collide/splash again
and eventually merge into one system. The molecular gas in the bridge
is on average forming stars at a rate of $\sim 6 \ms/yr$ 
($L_{\rm IR}=3.4\times 10^{10}$\ls). It has potential
for future starburst with ten-fold increase of $L_{\rm IR}$ 
in the next few tens  million years as 
the gas depletion time is more than 2 Gyr.


\subsection{Uniqueness of the Taffy Phenomenon and High-redshift 
Implications}

Recently a second taffy pair VV 769 = UGC 00813/6 was discovered 
(Condon, Helou \& Jarrett 2002), and the radio continuum 
and HI line images show many similarities
with that of UGC~12915/4. This is remarkable since such systems indeed
seem to be rare in the local universe (Condon \etal 2002). 
Nevertheless, systems with a radio continuum emission bridge between
the two galaxies do exist, e.g., Arp~270 
(NGC~3395/6) observed by  Huang et al. (1994), although they are
not as dramatic as the prominent taffy bridge in UGC~12915/4. 
On the other hand,
a substantial population of IR luminous galaxies 
with dust much cooler than the local LIGs
may exist at moderate/high redshifts (Chapman et al. 2002).
There is also some indication that there may be a high redshift population
of sub-mm sources with excess radio emission compared to the IR/radio
correlation for local galaxies.
Taffy does have excess radio continuum emission in the bridge and perhaps 
the conditions required to produce a taffy-type system
may have occurred more often 
at high redshift where collisions between
gas-rich systems are believed to be much more frequent. In particular,
its spectral energy distribution (SED) peaks beyond 100$\mu$m
suggesting a huge amount of cold dust in the Taffy system. In this
sense, it is crucial to obtain a highly spatially resolved 450$\mu$m
SCUBA map and future SIRTF 70 and 160$\mu$m maps in order to
fully investigate SED across different regions in the Taffy galaxies.
The study of the Taffy-like galaxies and better
characterization of their SED may thus be relevant to the high-z SCUBA sources.

It is also interesting to note that the target galaxy is a ring galaxy. 
Thus, the Taffy system is 
probably at least a remote relative of galaxies in 
the ``Sacred Mushroom'' class of ring galaxies (Arp \& Madore 1987). 
A Sacred Mushroom, e.g., AM 1724$-$622 (Wallin
\& Struck-Marcell 1994), is defined by a stem (intruder) consisting 
of an edge-on or disrupted companion connected to a ring galaxy 
forming the cap of the mushroom 
(target). An HI image of AM 1724$-$622 (also known as ESO 138 - IG 029/30)
clearly shows the HI 
bridge (Higdon \etal 2001). The intruder's orientation 
is different in the Taffy system as compared to the companion
of the Sacred Mushroom ring galaxies. The fact that the two galaxies 
are still in contact implies that the collision is not quite over.
Therefore, most of active star forming regions in the taffy bridge
must be fairly young and very recent.

Arp 284 (NGC 7714/5) may provide a better example. 
The recent high-resolution HI observations (Smith, 
Struck \& Pogge 1997) show that the bridge gas and HII regions are 
both spatially offset from the stellar bridge by a
large amount. According to models by Smith \& Wallin (1992) and  
Smith et al. (1997), the bridge may be the result of the combined 
action of gravitational (tidal bridges) and hydrodynamic (purely 
gaseous splash bridges) procesess. Most ring galaxies like
the Cartwheel and VIIZw 466 have gaseous splashes connecting
the companion to the target ring galaxy (Hidgon 1996; 
Appleton, Charmandaris \& Struck 1996). However, the amount of HI
in the bridge is small. 
The large amount of HI gas in the taffy bridge
shows how effective these processes can be in some cases.

Although the NRAO VLA sky survey (Condon \etal 1998) 
found only one other taffy 
system in the nearby universe, taffy phenomena exist in other 
forms of ISM besides relativistic particles responsible for radio continuum emission.
For instance, II~Zw~71 (UGC~9562) has been classified as a
``probable'' polar ring galaxy. However, it is connected with its 
companion galaxy II~Zw~70, a blue compact dwarf galaxy separated by a 
projected distance of 23 kpc, 
by a taffy-like HI bridge 
with no detectable radio continuum emission (Cox \etal 2001).
Duc et al. (2000) presented
another case (Arp~245) where two galaxies are connected in HI.
Quite a few more examples of HI bridges are listed in the
HI Rogues Gallery (Hibbard \etal 2001b), e.g., NGC~3424/30 
(Nordgren \etal 1997a) and NGC~7125/6 (Nordgren \etal 1997b). 
It is possible that weak extended radio continuum 
could exist in these taffy-like HI bridges, which requires 
much deeper and sensitive continuum imaging to probe. Nevertheless,
little star formation is occurring in most of the HI bridges
since apparently the collision damage to the molecular gas disks
is minimum and few GMCs were pulled 
out of the molecular gas disks. It is worth noting that none of
the examples mentioned are LIGs, although most have a large
gas reservoir (at least HI).

Other cases exist with radio continuum emission
outside of the galaxy disks, but not necessarily located 
exactly between the merging disks. In NGC~6670 (Wang et al. 2001), a huge
HI gas concentration is found between the two merging disks of
heavily distorted HI morphology. But the nearly edge-on 
merging geometry seems to produce an extra-disk HI overlap
which is not exactly between the nuclei. Weak radio continuum is also
apparent in the HI concentration outside of the disks of
NGC~6670. In the ``Antennae'' galaxies, the dominant radio 
continuum emission is also from the overlap region 
(Neff \& Ulvestad 2000; Gao et al. 2001a), which is not 
exactly between the two nuclei. II~Zw~96 also exhibits dominant 
radio continuum (Condon et al. 1996; Goldader et al. 1997) where 
the strongest CO emission is found entirely outside
of the two merging stellar disks, and also far away
from the region between the two galaxies (Gao et al. 2001b). It is
difficult to understand how collision/merging could 
produce so much gas and star formation outside of the
disks, and offset far away from the region between the two galaxies.
Nonetheless, these systems with extreme offsets in the gas 
overlap tend to be indeed bona-fide LIGs.

\section{CONCLUSIONS}

We presented two-field mosaicing CO images of 
the Taffy system, and compared the BIMA CO images
with various observations at other wavebands. In particular, 
we investigated the star formation efficiency (SFE) across the 
entire system, including the prominent taffy bridge, to identify
the starburst sites. The detailed molecular gas
distribution and kinematics provide useful constraints on
the orbital geometry and impact parameters of the 
interaction.  We summarize our main results as follows:

1. Large amounts of molecular gas were clearly detected throughout 
the taffy bridge between the two galaxies. The highest CO
concentration in the taffy bridge corresponds to the extra-disk
\ha~ source (HII region), also the strongest radio continuum source. 
The molecular gas mass in the taffy
bridge amounts to 1.4$\times 10^{10}\ms$, using the
standard Galactic CO-to-H$_2$ conversion.

2. The overall CO distribution of the entire 
system, including the taffy bridge, agrees well with that of the radio 
continuum emission, and argues for star formation 
as the origin of most of the radio continuum emission. 
We further show that the starburst sites are exclusively in
the intruder galaxy UGC~12915 and the adjacent extra-disk
\ha~ source in the taffy bridge. Most of the molecular
gas in the taffy bridge is forming into stars at 
SFE comparable to that of the target galaxy, and similar
to that in Galactic disk GMCs. These are derived based on the 
20cm-to-CO ratio (SFE) maps.

3. Compared to the HI morphology and kinematics, which are strongly
distorted owing to the high-speed collision, CO better defines
the orbital geometry and impact parameter of the interaction as
well as the disk properties. The two spiral disks are clearly 
counter-rotating at speeds above 300\kms. The intruder's major axis 
is more tilted relative to that of its companion than inferred from 
its optical appearance, which is heavily 
affected by the dust obscuration and tidal features.

4. CO observations and a multiwavelength comparison demonstrate
that the high-speed counter-rotating collision lead to the observed 
gas (mostly HI, some H$_2$) pile-up between the galaxies. The impact 
parameter must be
smaller than the disk radii so that the GMCs in the inner disks
and nuclear regions have had more chance to collide during
the interaction.

\acknowledgements

We thank Jim Condon, Tom Jarrett, and George Helou for providing
digital images and helpful discussions. YG was supported by the
NSF grant AST 01-00793 funded at the FCRAO. This research was supported in
part by a research grant to ERS from the Natural Sciences and Engineering
Research Council of Canada.

\clearpage

\begin{deluxetable}{lrrrrrr}
\tablenum{1}
\tablecolumns{7}
\tablecaption{Log of the BIMA CO(1-0) Observations in the Taffy Galaxies}
\tablehead{
\colhead{\#}        &     \colhead{Obs. Date}     &
\colhead{Array Config.}  &     \colhead{Track Duration}         &
\colhead{T$_{\rm sys}$}         &     \colhead{\# of Antennas}   &
\colhead{Grades\tablenotemark{a}}  \\
\colhead{   }        &     \colhead{1999}    &
\colhead{ }        &     \colhead{Hours}    &
\colhead{K}    &   \colhead{ }    &
\colhead{ } }

\startdata

  1 & Apr 27 & C & 7.9 & 290$\pm 70$ & 10 & B- \nl
  2 & Apr 29 & C & 8.0 & 310$\pm 70$ & 10 & B- \nl
  3 & May 6  & C & 4.0 & 330$\pm 60$ &  9 & C  \nl
  4 & Jun 5  & C & 9.0 & 400$\pm 110$ & 10 & B- \nl
  5 & Jun 14 & C & 6.7 & 300$\pm 70$ &  10& B  \nl
  6 & Jul 12 & D & 7.0 & 340$\pm 80$ &  7 & C  \nl
  7 & Jul 14 & D & 5.6 & 380$\pm 90$ &  7 & C  \nl
  8 & Jul 15 & D & 4.5 & 360$\pm 90$ & 10 & B  \nl
  9 & Jul 19 & D & 5.5 & 320$\pm 70$ &  9 & B  \nl
 10 & Jul 27 & D & 7.5 & 290$\pm 80$ & 10 & B+ \nl

\tablenotetext{a}{Overall grading based on system performance and weather
conditions etc. as given by duty observers.
A: Excellent, B: Good, C: OK/usable, D: Bad, F: Failure}

\enddata
\end{deluxetable}

\begin{deluxetable}{lrrrrrrr}
\tablenum{2}
\tablecolumns{8}
\tablecaption{CO Measurements of the Taffy Galaxies}
\tablehead{
\colhead{Source}        &     \colhead{$f_{\rm CO}$}     &
\colhead{$V_{\rm CO}$\tablenotemark{a}}  &     \colhead{M(H$_2$)}         &
\colhead{M(HI)}         &     \colhead{M$_{\rm dust}$}   &
\colhead{L$_{\rm IR}$\tablenotemark{b}}  &  \colhead{SFE\tablenotemark{c}}  \\
\colhead{   }        &     \colhead{Jy\kms}    &
\colhead{\kms }        &     \colhead{10$^{10}$\ms}    &
\colhead{10$^{10}$\ms}    &   \colhead{10$^{7}$\ms}    &
\colhead{10$^{10}$\ls}     &   \colhead{\ls/\ms}}

\startdata

  U12915 & 355 & 100 & 1.5 & 0.4 & 5. & 3.3 & 11 \nl
  U12914 & 315 & -60 & 1.3 & 0.5 & 6. & 1.4 & 4 \nl
  HII region &  90 & 90 & 0.4 & 0.06 & ... & ... & 7 \nl
  Bridge\tablenotemark{d} & 325 & 90 & 1.4 & 0.6 & ... & 3.4 & 4 \nl
  HI peak\tablenotemark{e} & $\approxlt 30$ & ... & 0.1 & 0.08 & ... & ... & 5 \nl

\tablenotetext{a}{Zero velocity corresponds to 4400\kms}
\tablenotetext{b}{Estimated by roughly scaling according to the 20cm 
radio continuum emission}
\tablenotetext{c}{Highest estimated from the SFE (=$L_{\rm IR}/M$(H$_2)$) 
map in Fig.~7}
\tablenotetext{d}{Entire bridge including the HII region and HI peak. But 
the highest SFE is referring to regions excluding the HII region and HI peak}
\tablenotetext{e}{HI peak region in a 18$''$ beam}

\enddata
\end{deluxetable}

\clearpage

\figcaption{CO(1-0) image of the Taffy system in false-color. The color
ranges linearly from 37 to 170 Jy\kms per synthesized beam.
The synthesized beam of $9.''9\times 9.''7$ is shown in the lower left. 
The white dashed
circles outline the two mosaicing fields mapped by the BIMA. \label{fig-1}} 

\figcaption{Velocity channel maps of 20\kms~ width in contours overlaid
on the DSS image. Top-left of each panel labels the velocity (here zero
corresponds 4400\kms). Contours start at 0.085 Jy/beam (nearly 4$\sigma$),
and increase successively by a factor of $\sqrt {2}$. The last panel is 
the velocity integrated (over entire 800\kms~ range) CO intensity 
map (moment zero map as in Fig.~1). Here contours start at 
32.5 Jy\kms/beam, and increase by a factor of $\sqrt {2}$. \label{fig-2}}

\figcaption{
VLA HI intensity contour map (Condon \etal 1993) overlaid
on DSS image to illustrate the various HI components. The lowest
contour corresponds to column density 5$\times 10^{19}$cm$^{-2}$. 
Successive contours 
increase by factors: 2, 4, 5.6, 8, 11.3, 16, 22.6, 32 (white), 
38.1, 45.2 (white), 50, 55, 60, 64 (white). \label{fig-3}}

\figcaption{(a) CO velocity field of the velocity integrated 
first moment map (zero corresponds 4400\kms); (b) CO velocity 
linewidth map (the velocity integrated second moment map). 
The color bars code the velocity scale in \kms. \label{fig-4}}

\figcaption{Position--Velocity maps. (a) Cut along the major axis
of the intruder UGC~12915 from SE to NW; (b) A slice joining the 
two nuclei across the bridge from NE to SW; (c) Cut along the 
major axis of the target UGC~12914 from SE to NW. Grey-scale image
and the white contours are HI of Condon et al. (1993) and the 
thick black contours are our convolved CO.
\label{fig-5}}

\figcaption{The DSS
contours are compared to the near-IR K$_s$ image (left) indicating
the underlying stellar disk is slightly less tilted. The dash box 
outlines the area in the bridge within which the $7\times 5$ grid 
spectra (right) at 9$''$ spacing for both CO (thin line) and 
HI (thick line) were extracted. The labels (shown in the lowleft 
spectra) are only for CO flux 
in Jy/beam, and velocity range is 3900 to 4860 km/s. \label{fig-6}} 

\figcaption{The 20cm-to-CO ratio contours overlaid on the CO image. 
The 20cm-to-CO ratio could be interpreted as star formation efficiency 
(SFE) if the 20cm continuum is indicative of the star formation rate. 
The first contour corresponds roughly to SFE$\sim 2 \ls/\ms$,
the rest increases successively by $\sqrt {2}$.
\label{fig-7}
}

\figcaption{CO contours compared to the multiwavelength images of
the near-IR H-band (a, topleft), ISO mid-IR 15$\mu$m (b, topright), 
VLA 20cm radio continuum (c, lowleft), and 21cm HI line (d, lowright). 
The CO contours of 22.5,
25, 27.5, 30, 35, 40, 50, 60, 70, 80, 90, 99\% of the peak 
emission are plotted in all panels. \label{fig-8}}

\end{document}